\documentclass[twocolumn]{aastex631}

\newcommand\aastex{AAS\TeX}

\usepackage{graphicx}

\begin{document}

\title{Template \aastex Article with Examples: 
v6.3.1\footnote{Released on March, 1st, 2021}}

\title{Exoplanet Detection by Machine Learning with Data Augmentation}

\author[0000-0002-9281-2684]{Koray Aydo\u{g}an} 
\affiliation{Department of Physics, Bo\u{g}azi\c{c}i University, Bebek 34342, \.{I}stanbul, Turkey}

\begin{abstract}
It has recently been demonstrated that deep learning has significant potential to automate
parts of the exoplanet detection pipeline using light curve data from satellites such as Kepler \cite{borucki2010kepler} \cite{koch2010kepler} and NASA's 
Transiting Exoplanet Survey Satellite (TESS) \cite{ricker2010transiting}.
Unfortunately, the smallness of the available datasets makes it difficult to realize the level of performance one expects 
from powerful  network architectures. 

In this paper, we investigate the use of data augmentation techniques on light curve data from
to train neural networks to identify exoplanets. The augmentation techniques used are of two 
classes: Simple (e.g. additive noise augmentation) and learning-based (e.g. first training a GAN \cite{goodfellow2020generative} to generate new examples). 
We demonstrate that data augmentation has a potential to improve model performance for the
exoplanet detection problem, and recommend the use of augmentation based on generative models as 
more data becomes available.

\end{abstract}

\section{Introduction} \label{sec:intro}
Attempts to discover and identify exoplanets by human eye is a very tough task to be solved due to the complexity of the light flux distribution on the light curve data. Moreover it is not feasible too because of the fact that the amount of the data which comes from the telescopes is so huge and performing the analysis is a very time consuming process. All of these factors lead to employing neural networks. 

Deep learning has been highly successful in a wide range of scientific problems where large
datasets are available. Unfortunately, in many fields, the lack of adequate data prevents one from obtaining
the full benefits of sophisticated deep learning models due to problems with overfitting.

One approach to dealing with this problem is to enlarge the existing datasets by using data 
augmentation techniques. Given a small dataset, some standard approaches to creating new, "augmented"
samples involve the transformation of existing samples in a way that is meaningful for the problem under
consideration  (e.g. rotating images) or by adding various forms of noise to existing samples. 

A more interesting approach to data augmentation involves the use of a machine learning model to generate new
samples by learning from existing samples. While this approach may also be prone to overfitting, 
it has long been recognized that it also has the potential to boost the performance of deep learning
models.

Exoplanet detection by light curve data is certainly a problem where datasets are not as large 
as one would like. For instance, the light curve dataset from  NASA's Transiting Exoplanet Survey Satellite (TESS)
includes 1 billion objects (TESS Input Catalog). While this number is steadily increasing, it is 
certainly much smaller than one would want when training a powerful deep learning model to detect exoplanet
candidates with a high degree of precision and recall. 

There are various works done by utilizing some deep learning methods for classification tasks with some specific datasets, which come from different telescopes, such as K2 photometry from Kepler Mission, \cite{vanderburg2014technique}; Kepler data from Kepler telescope \cite{shallue2018identifying}; and with TESS data from TESS satellite \cite{yu2019identifying}. 

In all of these works they followed a specific pipeline in order to prepare the whole dataset to be trained and tested. However, for the last two works, the whole preprocessing is the same; binning (splitting the data to global, local and secondary views), detrending processes can be given as examples. For the last two, the neural network architecture consists of 1-dimensional CNN’s \cite{o2015introduction} with different number of filter sizes. And, they got appreciable results according to precision and recall. However, for the first research, there is another work \cite{malik2022exoplanet} which employed a method, feature extraction, via a Python package TSFresh \cite{christ2018time}. And, by utilizing this process they obtained a simulated data, the rest is again machine learning. We need to note that they also applied the feature extraction process to the previous datasets (Kepler and TESS) and they got better results in metrics. 

Recently, there is another work done which uses CNNs again, but with a different number of input channels and different preprocessing algorithms \cite{valizadegan2022exominer}. The dataset that they used both TESS and Kepler datasets from Mikulski Archive for Space Telescopes (MAST). And, they got the best results up to now in each metric. They also validated 301 new exoplanets from MAST catalog.

Deep learning models have been utilized for this problem
with some success, but the  acquisition of larger amounts of data in the near future will certainly improve
the performance of such models.

In this paper, we explore the use of data augmentation techniques for expanding the light curve data 
obtained from the Transiting Exoplanet Survey Satellite (TESS) to train neural network classifiers
on larger, synthetically extended datasets.
The expansion of the natural dataset with synthetically augmented samples
improves the recall value obtained by the neural network classifier formulated by \cite{yu2019identifying}
from 0.57 to 0.67. 
While the existing dataset is too small to draw definitive conclusions, we find the results encouraging,
and are aiming to reiterate the process as additional data becomes available.

The organization of the paper is as follows. In Section 2, we provide a review of the data augmentation techniques
relevant for the current work. In Section 3, we describe
the dataset we use. In Section 4, we describe our augmentation approaches and the model architectures used, and
provide the results obtained. In Section 5, we explain the training process, and we provide conclusions and discuss possibilities for future work with Section 6.

\section{Data Augmentation Techniques}

Synthetically expanding an existing dataset in order to improve model performance is
a technique that has been in use for a long time in the machine learning literature.
One of the earlier examples is the use of "distorsions" (i.e. affine planar
transformations) to generate new images for the problem of handwritten character recognition.
A more recent prominent example is \cite{krizhevsky2017imagenet}, where augmentations such as 
image translations, reflections, and a form of scaling of the principal components were
used in an image classification problem.

By now, the use of data augmentation has a very large literature that is impossible to summarize here. For a broad overview of the use of data augmentation in image data we mention
\cite{shorten2019survey}. For a comparison of some data augmentation techniques on medial data, see
\cite{mikolajczyk2018data}. The data we use in this paper is of the form of time series; 
for a survey of data augmentation techniques in time series data,
see \cite{wen2020time}. 

Although it is hard to give a clean classification of data augmentation methods, we can mention
the following broad types of techniques:
\begin{itemize}
    \item Noise augmentation: Generating new samples from existing ones by incorporating some sort of noise.
    Perhaps the simplest example of this method is adding Gaussian noise to existing samples to create new ones.
    \item Data transformation: Applying forms of transformation to the data that preserve the "underlying meaning"
    that the model is trying to learn. For example, for a model doing object detection from images,
    rotating, flipping, or zooming into an image doesn't change the object content, so it makes sense 
    to create new examples by using these transformation.
    
    For time series data \cite{wen2020time}, depending on the problem, one may use techniques like shifting time, and
    time warping and amplitude warping, which
    non-uniformly stretch the time axis and the amplitude axis, respectively. However, note that only if the
    problem at hand is expected to have these sorts of transformations as invariances would it make sense to use
    these methods to create new samples.
    \item Using generative models: This approach involves training a generative model such as a generative adversarial network (GAN) \cite{goodfellow2020generative} on the existing data, and then using this model to create new, synthetic
    samples. While the transformation methods mentioned above are hand-picked, the generative approach delegates
    the creation of new realistic samples to machine learning, as well, without a need to search for meaningful 
    transformations. Of course, whether this approach actually works well would depend on the problem at hand.
\end{itemize}

There is a large literature on the use of GANs for data augmentation alone. See \cite{yi2019generative} for examples of
this approach in medical imaging, and \cite{shorten2019survey} where such techniques are discussed 
as par tof a more general survey. We can also mention \cite{antoniou2017data} as a novel specific
technique that uses a slightly modified form of GAN to use with data augmentation.

In this paper, we will use the following augmentation techniques in TESS light curve data,
which comes in the form of three time series per observation:
\begin{itemize}
    \item Noise augmentation
    \item GAN-based augmentation
\end{itemize}

For noise augmentation, we create new samples by adding noise generated from various distributions
such as the normal distribution. For GAN-based augmentation, we train separate GANs on "planet" samples
and "non-planet: samples.

\section{Data}
In our research, we chose a light curve dataset that is obtained from the Transiting Exoplanet Survey Satellite (TESS), the pipeline is done by \cite{yu2019identifying}. The main reason why we chose this dataset is the number of exoplanets in the whole dataset is low (492 exoplanets out of 15959 data samples), the dataset is also small. This dataset includes information about brightness, time, moment-derived centroid and so forth. While we are training, we consider only information about the brightness, we consider “global view”, “local view”, and “secondary eclipse” keys in the whole dataset, like did in the previous work. “global view”, which shows the light curve over an entire orbital period; and a “local view”, which is a close-up of the transit event and for the secondary eclipse what we observe is that when the planet crosses behind the star or the galaxy the light from the planet is blocked and we see the decrease in the flux. Furthermore, global view has 201 data points the rest have 61 data points \cite{yu2019identifying}.

\begin{figure}[h!]
  \includegraphics[width=9cm]{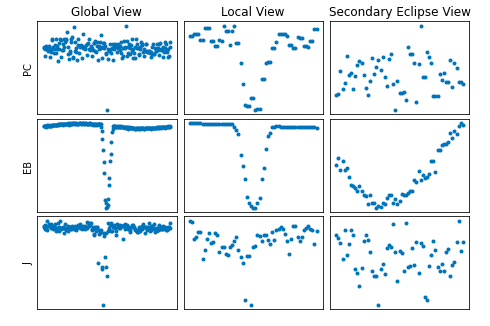}
  \caption{Light Curve Data Samples. PC (Planet Candidate); EB (Eclipsing Binary); J (Junk) with their Global, Local and Secondary views}
\end{figure}

\section{Our Approaches}
\subsection{Classical Approach (Noise Augmentation)}

As we discussed some of the data augmentation techniques in the previous section, it is a common technique employed in deep learning in order to improve the performance and as we said earlier, we used noise augmentation technique for this work. 

Noise augmentation is a very common method that is applied for both image based data and time series data, seismological data can be given as an example \cite{mousavi2020earthquake}. For this example noise augmentation is done with only Gaussian probability density function. In the line with that density function, we also used different probabilistic density functions which are, exponential and Rayleigh probabilistic density functions in two approaches; we added noise to the training set and we multiplied the noise with the training set. Then, we concatenate the augmented data to the original data, so we got expanded version of the training dataset.

\subsection{Deep Learning Based Approach (GANs)}

As discussed in the early section, machine learning based data augmentation is another common and new method to expand a training dataset. Its architecture is similar to autoencoders, in that case we had encoder and decoder parts, now we have two models which are called “generator” and “discriminator”. Generator part creates a random seed, the discriminator part classifies whether the data is fake or real, and the training process continues up to an equilibrium point. In our work, we produced synthetic exoplanet data and noise with a GANs architecture. Then, as we did in noise augmentation, we concatenate the synthetic data to the original data and trained the deep learning model. The main difference from the previous works such as \cite{yi2019generative} is we produced one dimensional data, time series data instead of an image dataset.

\begin{figure}[h!]
  \includegraphics[width=9cm]{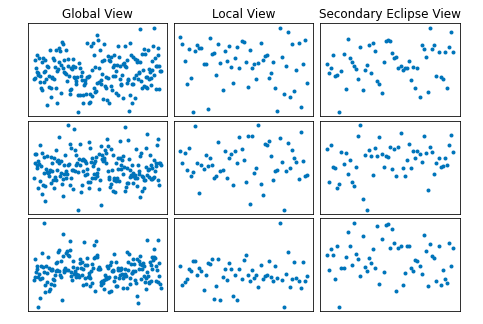}
  \caption{Synthetic samples of Global, Local and Secondary Eclipse Views produced by our GAN architecture}
\end{figure}

Our GANs architecture consists of several layers of one-dimensional CNNs whose generator part has three output channels whereas the discriminator has 3 input channels to be compatible with global, local and secondary view channels.

\section{Training}

As said earlier, we employed noise augmentation technique in two approaches; additive and multiplicative noise. We started with Gaussian density function and we augmented the raw training data with Gaussian noise. We did the same process for the other density functions i.e. exponential and Rayleigh. With the augmented data, we trained it with the same model \cite{yu2019identifying}.

The determination of the scale parameters of these functions is based on visual check. In other words, we specified the parameters according to the visual forms the augmented data; we did not deform the data extremely. 

For the deep learning based method, we expanded the training dataset with two different ways;

\begin{itemize}
    \item Producing Synthetic Exoplanet (Signal) Dataset 
    \item Producing Noise (Non-Signal) Dataset
\end{itemize}

For the first method, we separated the exoplanets ("PC" labeled data points) from the training dataset and fed them into our GANs model. And, we ended up with different numbers of synthetic planet candidates that we specified. In this approach, we produced 1000, 2000, 4000 and 6000 synthetic signals. 

We did this up to 6000 in order to obtain a homogeneous training dataset since the whole training dataset consists of approximately 13000 data points. 

After the production phase, we expanded the original training dataset with these synthetic datasets gradually and trained the model \cite{yu2019identifying} with the same parameters that is specified.

For the second method, we cut off the rest of the exoplanets (non-"PC" labeled data points) from the training data and again we trained our GANs architecture with these noise data in order to produce synthetic noise. We produced noise with the same number of training dataset size to duplicate the data. 

In both circumstances, we trained our GANs architecture with 200 epochs, 256 batch size and Adam optimizer \cite{kingma2014adam}. Also, our GANs architecture is made of Convolutional Neural Networs, by using a deep learning package which is TensorFlow \cite{tensorflow2015-whitepaper}.

\section{Results}

In the line with our methods of data augmentation techniques, we got better recall score compared to the previous work \cite{yu2019identifying} when we produced synthetic planet candidate light curve data points and expanded the training dataset with them.

\begin{figure}[h!]
  \includegraphics[width=9cm]{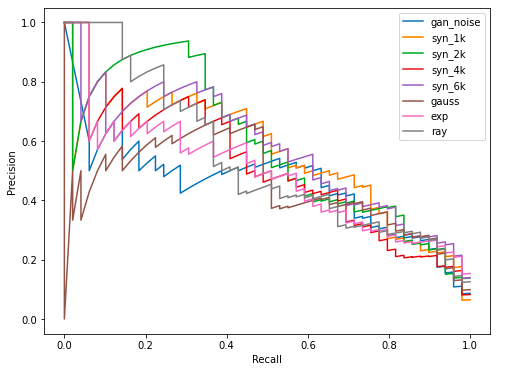}
  \caption{Precision-Recall Curve of Each Augmentation Method}
\end{figure}

The above plot is created for $gan noise$ we expand the training set by adding synthetic noise that is created by our GANs architectures. For $syn 1k$, $syn 2k$, $syn 4k$, $syn 6$, we expand the training set by concatenating synthetic planet candidate light curve signals to the original training set. For $gauss$, $exp$, $ray$, again we expand the training dataset by adding Gaussian noise, exponential noise and Rayleigh noise respectively. 

According to above augmentation methods, the architecture can distinguish Planet Candidates, PCs, from Eclipsing Binaries, EBs in a more successful way compared to the previous work, \cite{yu2019identifying} at a threshold of 0.5; in their work, they can recover 28 PCs out of 49, 0.67 recall, on the test set at the same threshold. We can recover 33 PCs out of 49, 0.67 recall, on the test set when we expanded our training dataset by concatenating 6000 synthetic exoplanets. The second successful result is 31 PCs out of 49 when we augmented the training dataset by expanding the training dataset with 4000 synthetic exoplanets that is created by our GANs archtitecture. 

\begin{figure}[h!]
  \includegraphics[width=9cm]{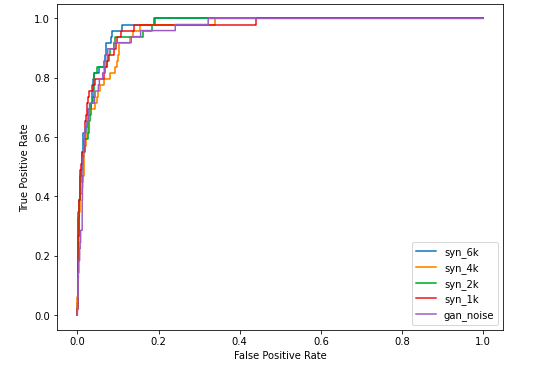}
  \caption{The ROC Curve for GANs Based Augmentation Method}
\end{figure}

For the AUC score metric for ROC curve, we got similar results with the previous work. In the previous work, they got 0.978 AUC score and our average AUC score is 0.970.

\section{Future Work and Conclusions}

To sum up, data augmentation techniques are developing and have a great impact on machine learning problems. In lots of deep learning research, conventional techniques are used, which are flipping, rotating, adding noise and so forth. But, in recent years, deep learning based data augmentation techniques are also used and their effects are so considerable. In our research, we applied both techniques and we got better recall score compared to the previous work and similar AUC score for the ROC curve by expanding our training dataset with planet candidate light curve signals. In the future, we think of applying these techniques to the most recent work, ExoMiner \cite{valizadegan2022exominer}, and compare the final prediction results.

Another approach that we are considering is that, performing transfer learning from the previous work for Kepler dataset \cite{shallue2018identifying} to this work. In other words, we will train the model partially on Kepler dataset and and continue with TESS dataset. 

\section{Data Availability and Code}

The synthetic data and the codes in this article will be shared on reasonable request to the corresponding author.

\bibliography{sample631}{}

\begin{thebibliography}{}
\expandafter\ifx\csname natexlab\endcsname\relax\def\natexlab#1{#1}\fi
\providecommand{\url}[1]{\href{#1}{#1}}
\providecommand{\dodoi}[1]{doi:~\href{http://doi.org/#1}{\nolinkurl{#1}}}
\providecommand{\doeprint}[1]{\href{http://ascl.net/#1}{\nolinkurl{http://ascl.net/#1}}}
\providecommand{\doarXiv}[1]{\href{https://arxiv.org/abs/#1}{\nolinkurl{https://arxiv.org/abs/#1}}}

\bibitem[{Abadi {et~al.}(2015)Abadi, Agarwal, Barham, Brevdo, Chen, Citro,
  Corrado, Davis, Dean, Devin, Ghemawat, Goodfellow, Harp, Irving, Isard, Jia,
  Jozefowicz, Kaiser, Kudlur, Levenberg, Man\'{e}, Monga, Moore, Murray, Olah,
  Schuster, Shlens, Steiner, Sutskever, Talwar, Tucker, Vanhoucke, Vasudevan,
  Vi\'{e}gas, Vinyals, Warden, Wattenberg, Wicke, Yu, \&
  Zheng}]{tensorflow2015-whitepaper}
Abadi, M., Agarwal, A., Barham, P., {et~al.} 2015, {TensorFlow}: Large-Scale
  Machine Learning on Heterogeneous Systems.
\newblock \url{https://www.tensorflow.org/}

\bibitem[{Antoniou {et~al.}(2017)Antoniou, Storkey, \&
  Edwards}]{antoniou2017data}
Antoniou, A., Storkey, A., \& Edwards, H. 2017, arXiv preprint arXiv:1711.04340

\bibitem[{Borucki {et~al.}(2010)Borucki, Koch, Basri, Batalha, Brown, Caldwell,
  Caldwell, Christensen-Dalsgaard, Cochran, DeVore,
  {et~al.}}]{borucki2010kepler}
Borucki, W.~J., Koch, D., Basri, G., {et~al.} 2010, Science, 327, 977

\bibitem[{Christ {et~al.}(2018)Christ, Braun, Neuffer, \&
  Kempa-Liehr}]{christ2018time}
Christ, M., Braun, N., Neuffer, J., \& Kempa-Liehr, A.~W. 2018, Neurocomputing,
  307, 72

\bibitem[{Goodfellow {et~al.}(2020)Goodfellow, Pouget-Abadie, Mirza, Xu,
  Warde-Farley, Ozair, Courville, \& Bengio}]{goodfellow2020generative}
Goodfellow, I., Pouget-Abadie, J., Mirza, M., {et~al.} 2020, Communications of
  the ACM, 63, 139

\bibitem[{Kingma \& Ba(2014)}]{kingma2014adam}
Kingma, D.~P., \& Ba, J. 2014, arXiv preprint arXiv:1412.6980

\bibitem[{Koch {et~al.}(2010)Koch, Borucki, Basri, Batalha, Brown, Caldwell,
  Christensen-Dalsgaard, Cochran, DeVore, Dunham, {et~al.}}]{koch2010kepler}
Koch, D.~G., Borucki, W.~J., Basri, G., {et~al.} 2010, The Astrophysical
  Journal Letters, 713, L79

\bibitem[{Krizhevsky {et~al.}(2017)Krizhevsky, Sutskever, \&
  Hinton}]{krizhevsky2017imagenet}
Krizhevsky, A., Sutskever, I., \& Hinton, G.~E. 2017, Communications of the
  ACM, 60, 84

\bibitem[{Malik {et~al.}(2022)Malik, Moster, \& Obermeier}]{malik2022exoplanet}
Malik, A., Moster, B.~P., \& Obermeier, C. 2022, Monthly Notices of the Royal
  Astronomical Society, 513, 5505

\bibitem[{Miko{\l}ajczyk \& Grochowski(2018)}]{mikolajczyk2018data}
Miko{\l}ajczyk, A., \& Grochowski, M. 2018, in 2018 international
  interdisciplinary PhD workshop (IIPhDW), IEEE, 117--122

\bibitem[{Mousavi {et~al.}(2020)Mousavi, Ellsworth, Zhu, Chuang, \&
  Beroza}]{mousavi2020earthquake}
Mousavi, S.~M., Ellsworth, W.~L., Zhu, W., Chuang, L.~Y., \& Beroza, G.~C.
  2020, Nature communications, 11, 1

\bibitem[{O'Shea \& Nash(2015)}]{o2015introduction}
O'Shea, K., \& Nash, R. 2015, arXiv preprint arXiv:1511.08458

\bibitem[{Ricker {et~al.}(2010)Ricker, Latham, Vanderspek, Ennico, Bakos,
  Brown, Burgasser, Charbonneau, Clampin, Deming,
  {et~al.}}]{ricker2010transiting}
Ricker, G.~R., Latham, D., Vanderspek, R., {et~al.} 2010, in American
  Astronomical Society Meeting Abstracts\# 215, Vol. 215, 450--06

\bibitem[{Shallue \& Vanderburg(2018)}]{shallue2018identifying}
Shallue, C.~J., \& Vanderburg, A. 2018, The Astronomical Journal, 155, 94

\bibitem[{Shorten \& Khoshgoftaar(2019)}]{shorten2019survey}
Shorten, C., \& Khoshgoftaar, T.~M. 2019, Journal of big data, 6, 1

\bibitem[{Valizadegan {et~al.}(2022)Valizadegan, Martinho, Wilkens, Jenkins,
  Smith, Caldwell, Twicken, Gerum, Walia, Hausknecht,
  {et~al.}}]{valizadegan2022exominer}
Valizadegan, H., Martinho, M.~J., Wilkens, L.~S., {et~al.} 2022, The
  Astrophysical Journal, 926, 120

\bibitem[{Vanderburg \& Johnson(2014)}]{vanderburg2014technique}
Vanderburg, A., \& Johnson, J.~A. 2014, Publications of the Astronomical
  Society of the Pacific, 126, 948

\bibitem[{Wen {et~al.}(2020)Wen, Sun, Yang, Song, Gao, Wang, \&
  Xu}]{wen2020time}
Wen, Q., Sun, L., Yang, F., {et~al.} 2020, arXiv preprint arXiv:2002.12478

\bibitem[{Yi {et~al.}(2019)Yi, Walia, \& Babyn}]{yi2019generative}
Yi, X., Walia, E., \& Babyn, P. 2019, Medical image analysis, 58, 101552

\bibitem[{Yu {et~al.}(2019)Yu, Vanderburg, Huang, Shallue, Crossfield, Gaudi,
  Daylan, Dattilo, Armstrong, Ricker, {et~al.}}]{yu2019identifying}
Yu, L., Vanderburg, A., Huang, C., {et~al.} 2019, The Astronomical Journal,
  158, 25

\end{thebibliography}
\bibliographystyle{aasjournal}

\end{document}